\documentclass[prl,showpacs,twocolumn,floatfix]{revtex4} 
\usepackage{epsfig} 
\usepackage{mathrsfs} 
\usepackage{amssymb} 
\usepackage{color} 
\usepackage{graphics}
\usepackage{psfrag}
\usepackage{graphicx} 
\newcommand{\ecen}{\end{center}} \newcommand{\btab}{\begin{tabular}} 
\newcommand{\etab}{\end{tabular}} \newcommand{\bdes}{\begin{description}} 
\newcommand{\edes}{\end{description}}  
 \newcommand{\beq}{\begin{equation}} 
\newcommand{\eeq}{\end{equation}} \newcommand{\bea}{\begin{eqnarray}} 
\newcommand{\eea}{\end{eqnarray}}  
  
\newcommand{\bary}{\begin{array}} \newcommand{\eary}{\end{array}} 
\newcommand{\benum}{\begin{enumerate}} 
\newcommand{\eenum}{\end{enumerate}} \newcommand{\bitem}{\begin{itemize}} 
\newcommand{\eitem}{\end{itemize}} % %bold greek characters % 

\begin{document}

%\preprint{MRI-P-011004}

%\title{Bulk Metal Insulator transitions in the Metal-Insulator-Metal sandwich}
\title{Physics of interface: Barrier with correlations and disorder 
sandwiched between two metallic planes}


\author{Sanjay Gupta} \email[]{sanjay1@bose.res.in}
\affiliation{ S. N. Bose National Centre for Basic Sciences, Kolkata, India}
\author{Tribikram Gupta } \email[]{trbkrm01@yahoo.co.in}
\affiliation{Theoretical Physics Division, Indian association for the Cultivation of Sciences, Kolkata, India}
%\thanks{}

%\collaboration{Sanjay Gupta}
%\email[]{sanjay1@bose.res.in}
%\affiliation{ S. N. Bose National Centre for Basic Sciences, Kolkata, India}

\date{\today}

\begin{abstract}
 A Metal-Disordered Mott insulator-Metal heterostructure is studied at half-fiiling using 
unrestricted Hartree Fock method. 
The corresponding clean system has been shown to be an insulator for any finite
on site correlation. Interestingly we find that introduction of explicit disorder
 induces a metal-insulator transition at a critical value of disorder. The
critical value  corresponds to the point at which disorder nullifies the effect of 
onsite correlation. The wavefunctions are found to delocalize by increasing disorder, 
rendering the system metallic.

\end{abstract}

%\pacs{73.20.-r, 71.27.+a, 71.30.+h,   , 71.10.Fd, 73.21.b, 73.40.c}
%\keywords{}

\maketitle

\section{Introduction} 
The field of hetterostructures, has seen tremendous interest in recent times
both experimentally\cite{Ohtomo, Thiel, Kenji} and theoretically\cite{Lee,Okamoto,Rosch, Krish,Gupta2}. 
Heterostructures fabricated out of strongly correlated systems, in particular have shown 
very novel properties. These emergent properties are sometimes very different from the 
properties of the individual constituent materials. For example the heterostructure
of a Mott insulator and a band insulator has been shown to have metallic behaviour\cite{Ohtomo}.
This is due to aggregation of delocalized electrons at the interface of such a heterostructure.
On lowering the temperature, the system can even become superconducting\cite{Thiel}. 
 
In this paper, we investigate a metal-disordered Mott insulator heterostructure. The corresponding 
clean case system has recently received lot of theoretical attention and has been variously claimed 
to be a fragile Fermi liquid\cite{Krish} which shows metallic character at any finite correlation 
strength, and a "de facto impenetrable Mott insulator" \cite{Rosch}, with highly decaying spectral 
weight at the Fermi energy. In a recent work we\cite{Gupta2} have shown that the clean system is 
insulating for all finite values of correlation that we could work with. The system was also found
to have multiple gaps in the energy spectrum, as a result of charge reconstruction on varying correlations.
  
An interesting relevant question in this context is the role of disorder on the whole phenomenon. 
Experimentally heterostructures are fabricated with a certain degree of disorder present in 
them. The disorder can arise dueto non trivial local atomic and electronic structure due to the 
presence of dangling bonds and incomplete atomic coordination. With these fundamental issues in 
mind, we have investigated a heterostructure comprising of a metal-disordered Mott insulator-metal 
sandwich. Very interestingly, we find that the introduction of disorder can generate a metal insulator 
transition in the region where the corresponding clean system was gapped. This is because disorder is 
able to nullify the effect of correlation in a certain range of values and also delocalize the wave 
functions. The route to metallicity has very close connection to the metallic phase 
that is stabilised by increasing disorder from a clean Mott insulator, due to the closing 
of the Mott gap as shown by Aguiar et. al \cite{Aguiar}, using generalized DMFT methods.  

In the system we study here, there is charge reconstruction between the planes 
and consequent inhomogenity in the underlying charge landscape. This inhomogenity was 
found to be sufficient to render the system insulating even in the gapless state of the clean 
system. Here we show that disorder is able to negate the effect of correlation over a small range of 
values and generates a metallic phase.   

We work with correlated disorder of the Thue-Morse and Fibonacci type.
Correlated disorder has been studied extensively before\cite{Merlin,Delahaye,Sanjay1,Roati}. 
Theoretical studies of the quasi-periodic systems have shown that in the
non-interacting limit the wave function shows a power law localization both in one\cite{Kohomoto} 
and two\cite{Sutherland} dimensions. A very recent experiment in a 1D quasi-periodic optical 
lattice\cite{Roati}, where the system was described by an Aubry-Andre Hamiltonian, has shown 
exponentially localized states(Anderson localization) in the large disorder limit. A theoretical 
explanantion of the above result has been attempted using Fibonacci sequence in 1D \cite{Modguno}    

Correlated disorder introduces a certain type of disorder which 
does not depend on the realisation, but still breaks translational sysmmetry.
Its relevance in capturing the effect of interplay between correlation and 
disorder has been shown recently by us\cite{Gupta1}. The two chosen sequences 
have been wrapped over the quasi 2D system in the Mott layers. 
%The similarity of our results qualitatively with a similar work done by Dariush et. al. 
%\cite{Nandini}, demonstrated the utility of this form of disorder. 

We use unrestricted Hartree Fock to study the problem as it is able to capture the 
role of disorder exactly and the effect of spatial variation in the order parameters 
both in plane and perpendicular to them. The effect of correlation is captured 
in a fully self consistent manner. It works very well in higher dimensions at half 
filling and even in 2D \cite{Hirsch} and 1D \cite{Sanjay} gives results comparable to 
QMC and real space renormalization calculations. 
It cannot however capture the effect of large effective mass renormalization and the 
evolution of the sharp metallic peak in the spectral function at the Fermi energy.  

%\section{A brief qualitative discussion}
%\section{Analytic Details}
%\section{Details of our calculation.}

\subsection{Model and Method} We now present and discuss our work. We take 
the barrier of our metal-barrier-metal sandwich to be described by the 
single-orbital disordered Hubbard model, and the metallic planes by the 
tight-binding model.  The Hamiltonian for the system is 

\begin{eqnarray} 
\mathcal{H}&=&\sum_{i\sigma}(\epsilon_{i} - \mu)n_{i\sigma}-\sum_{ij\alpha\sigma}t_{ij} 
c^\dagger_{i\sigma}c^{}_{j\sigma} \nonumber\\
&+& \sum_{i\alpha}U_\alpha(n_{i\alpha\uparrow} 
-\frac{1}{2})(n_{i\alpha\downarrow}-\frac{1}{2}).\; 
\end{eqnarray} 

Here $i,j$ is the site index.The operator 
$c^\dagger_{i\alpha\sigma}$ ($c^{}_{i\sigma}$) creates (destroys) an 
electron of spin $\sigma$ at site $i$. We set both the 
in-plane and the inter plane hopping to be nearest neighbor only, and equal to 
$t$. The lattice structure is that of a 
simple cubic lattice. $\epsilon_{i}$ denotes the site potential at the ith site which
follows either a Thue-Morse or a Fibonacci sequence in the barrier planes and are zero
in the metallic planes.  

A Thue-Morse sequence is generated by following the rule:
$A \rightarrow AB$ and $B \rightarrow BA$, where $A$ and $B$ are the two different sites.
In our case site $B$ has a higher site potential than site $A$ ($\epsilon_B > \epsilon_A$,
$W=\epsilon_B-\epsilon_A$, being the disorder strength). Thue-Morse chain in a particular 
generation looks like: $ABBABAABBAABABBA..$. 
The $A$ and $B$ sites are found in equal numbers in any Thue Morse sequence of even entries. 
The generating scheme of Fibonacci sequence is $A \rightarrow AB$ and $B \rightarrow A$.
 A typical Fibonacci chain in a particular generation looks
like: $ABAABABAABAABABAABABA......$. Here sites $A$ and $B$ follows a golden mean ratio. 
These two sequences correspond to two different impurity site concentration in the system.
In Thue Morse 50\% of the barrier sites are $B$ type, while in Fibonacci the corresponding 
number is 38\% . It is therefore expected that the effect of disorder
would be more strong for the Thue Morse disorder compared to the Fibonacci disorder.  
These sequences have been wrapped over the quasi 2D lattice in the barrier planes.
The interaction term is written in a particle-hole symmetric fashion. 
The chemical potential $\mu$ is calculated by demanding 
that there be exactly N electrons in the problem and is calculated by taking the 
average of the N/2 th and the (N/2 + 1) th energy level. We label the planes by $\alpha$ 
with $\alpha = m$ or $\alpha = 1$ corresponding to the 
metallic layers. All the other $\alpha$ values in between correspond to the 
barrier planes. Thus $U_1$ = $U_m$ = 0, while for all other $\alpha$, $U_{\alpha}$ = $U$.

\begin{figure}
\begin{center}
   \begin{tabular}{cc}
      \resizebox{39.5mm}{!}{\includegraphics{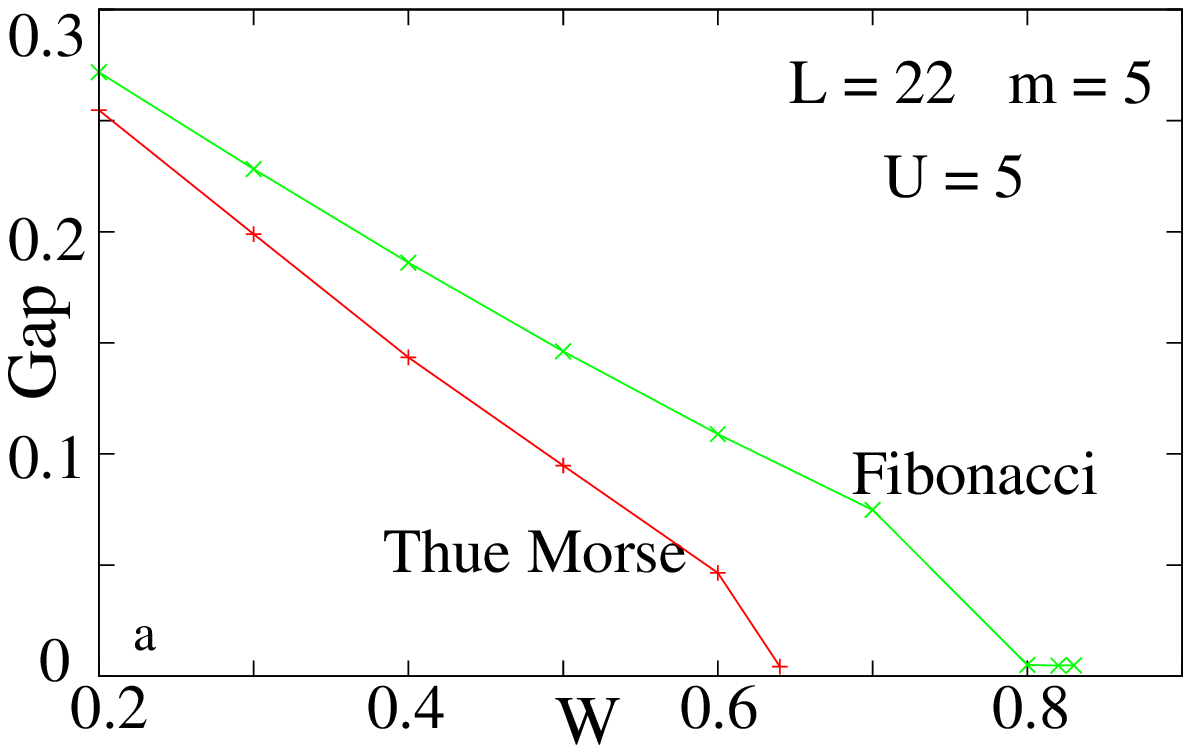}} &
      \resizebox{39.5mm}{!}{\includegraphics{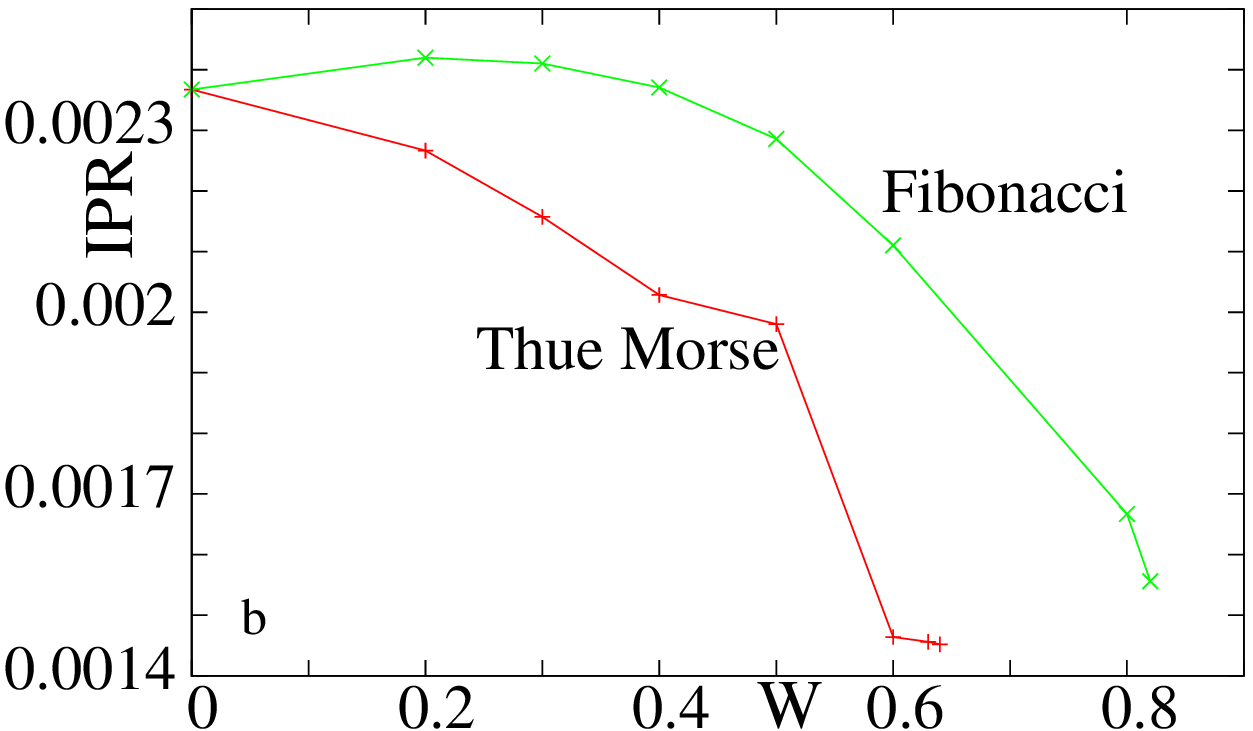}} \\
    \end{tabular}
\caption{a. Variation of gap for $U$ = 5 with increasing disorder for Thue Morse and Fibonacci sequences.
b. IPR with increasing disorder for $U = 5$ for Fibonacci and Thue Morse sequences}
\label{fig;Spin_ChargeOrder}
\end{center}
\end{figure}

In a recent work \cite{Gupta2} we had studied the interface of a quasi 2D barrier with 
onsite correlations sandwiched between two metallic planes and shown that 
even a very small correlation stength renders the system insulating. The 
system for very low $U$ behaves like a gapless insulator, while for larger 
$U$, the system opens a gap at half filling and then behaves like a gapped 
insulator. For even higher $U$, the system shows the presence of multiple 
gaps. In this work we will fix $U$ to be in the region where the system has opened 
a single gap. We then turn on the disorder strength from zero
and study the evolution of this sytem. 

We use unrestricted Hartree Fock method to address the problem as we do not want to 
restrict or bias the solutions in any particular manner. The price we pay in the process is 
the time taken to attain self consistency in the solutions, upto the high precision 
that we seek. This precision value is $10^{-7}$ in the square of the density of both 
up and down electron density. We need to perform about 100 iterations away from the 
transition point to reach self consistency, while close to the metal-insulator transition 
point we need to perform more than 1000 iterations to reach the same consistency. For 
disorder strength even higher than the transition value, it is very difficult to reach 
self consistency as the numerical solution vacillates between several close lying minima 
solutions. Hence we avoid this region of parameters in this work. The initial seed that 
we provide to the system becomes increasingly important as the system approaches the 
metal insulator transition point. Deep in the Mott insulator regime we find that 
though the system is not highly sensitive to the initial seed in determining the 
outcome of the final ground state, a mixed antiferromagnetic seed gives the 
lowest energy for both Fibonacci and Thue Morse type of disorder. 
Close to the transition point we find the initial seed that gives the 
lowest energy is different for the two types of disorder. This is because 
close to the transition point the effective single particle potential landscape 
that emerges due to competition between correlation and disorder becomes highly 
sensitive to the concentration and distribution of the impurity sites.   

\subsection{Calculated quantities} 
We study the effect of increasing $W$ on the gap at half filling, 
the spin and charge profile and the Kubo 
conductivity at zero temperature. 
The charge at a particular site is simply calculated as 
$C$ = $n_{i,\uparrow}$ + $n_{i,\downarrow}$. The spin at a 
particular site is given by $S$ = $|(n_{i,\uparrow} - n_{i,\downarrow})|$.
The optical conductivity is calculated using the 
Kubo formula, which at any temperature is given by:
\begin{equation}
\label{eq.8}
	\sigma ( \omega)
	= {A \over N}
	\sum_{\alpha, \beta}(n_{\alpha}-n_{\beta})
	{ {\vert f_{\alpha \beta} \vert^2} \over {\epsilon_{\beta}
	- \epsilon_{\alpha}}}
	\delta(\omega - (\epsilon_{\beta} - \epsilon_{\alpha}))
\end{equation}
%with $A = {\pi  e^2 }/{{h a_0}}$, $a_0$ being the lattice spacing,
with $A = {\pi  e^2 }/{{\hbar a_0}}$, $a_0$ being the lattice spacing,
and $n_{\alpha}$ = Fermi function with energy $\epsilon_{\alpha}-\mu$.
The $f_{\alpha \beta}$ are matrix elements of the current operator
 $j_x = i t  \sum_{it, \sigma} (c^{\dagger}_{{i + x a_0},\sigma}
 c_{i, \sigma} - h.c)$, between exact single particle eigenstates
 $\vert \psi_{\alpha}\rangle$,
 $\vert \psi_{\beta}\rangle$, {\it  etc},  and
 $\epsilon_{\alpha}$, $\epsilon_{\beta}$
 are the corresponding eigenvalues. In this paper, conductivity/conductance
 is expressed in units of $A = {\pi  e^2 }/{{\hbar a_0}}$.

We calculate the `average' conductivity over 3-4 small frequency intervals 
$\Delta \omega$($\Delta \omega$ = $n \omega_{r}$, n = 1,2,3,4),
and then differentiate the integrated conductivity to get $\sigma(\omega)$
at $\omega = n \omega_{r}$, n = 1,2,3,4\cite{SanjeevEPL,Gupta1}. The $\omega_r$
is taken to be twice the lowest frequency that can be accessed of a finite size 
lattice($\omega_r = 2*D/N$), where $D$ is the bandwidth and N is the total number 
of sites in the system. 

For the quasi 2D geometry taken by us $N = m\times{L^2}$, where $m$ is the total 
number of layers (metal + insulator) and L is the total number of sites along 
both x and y direction, in the numerical work. 

\begin{figure}
\begin{center}
   \begin{tabular}{cc}
      \resizebox{39.5mm}{!}{\includegraphics{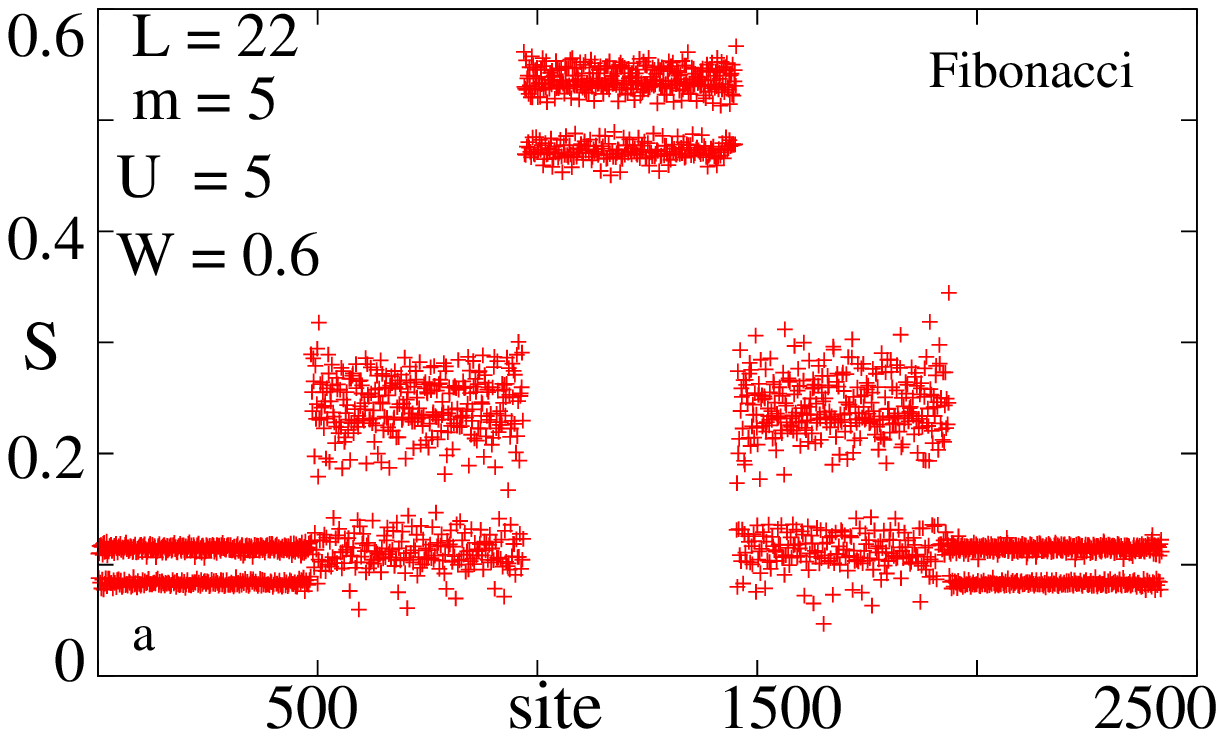}} &
      \resizebox{39.5mm}{!}{\includegraphics{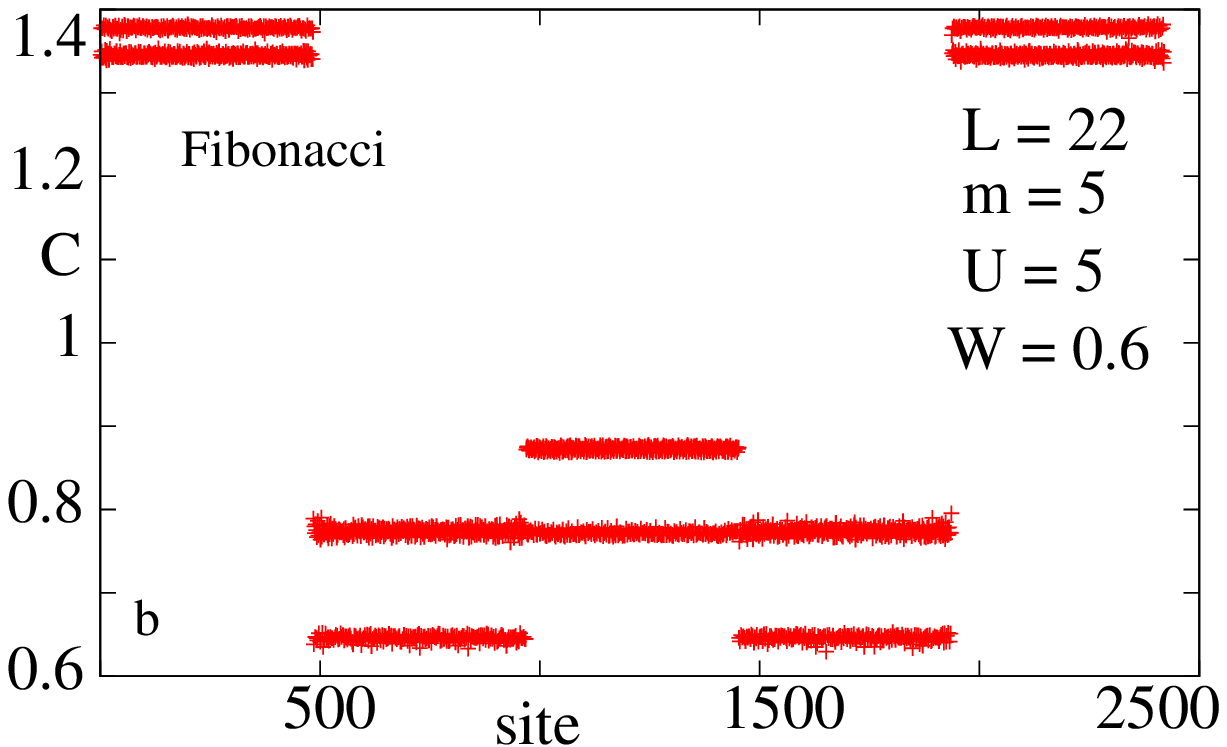}} \\
      \resizebox{39.5mm}{!}{\includegraphics{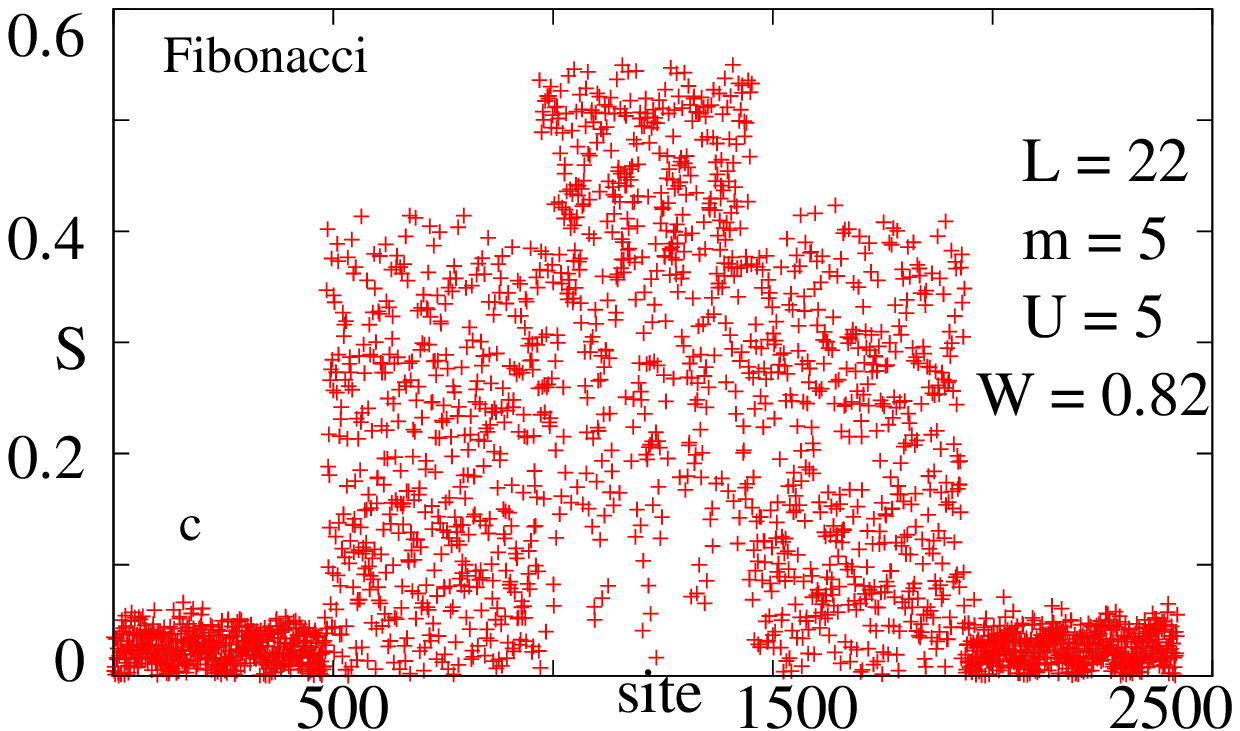}} &
      \resizebox{39.5mm}{!}{\includegraphics{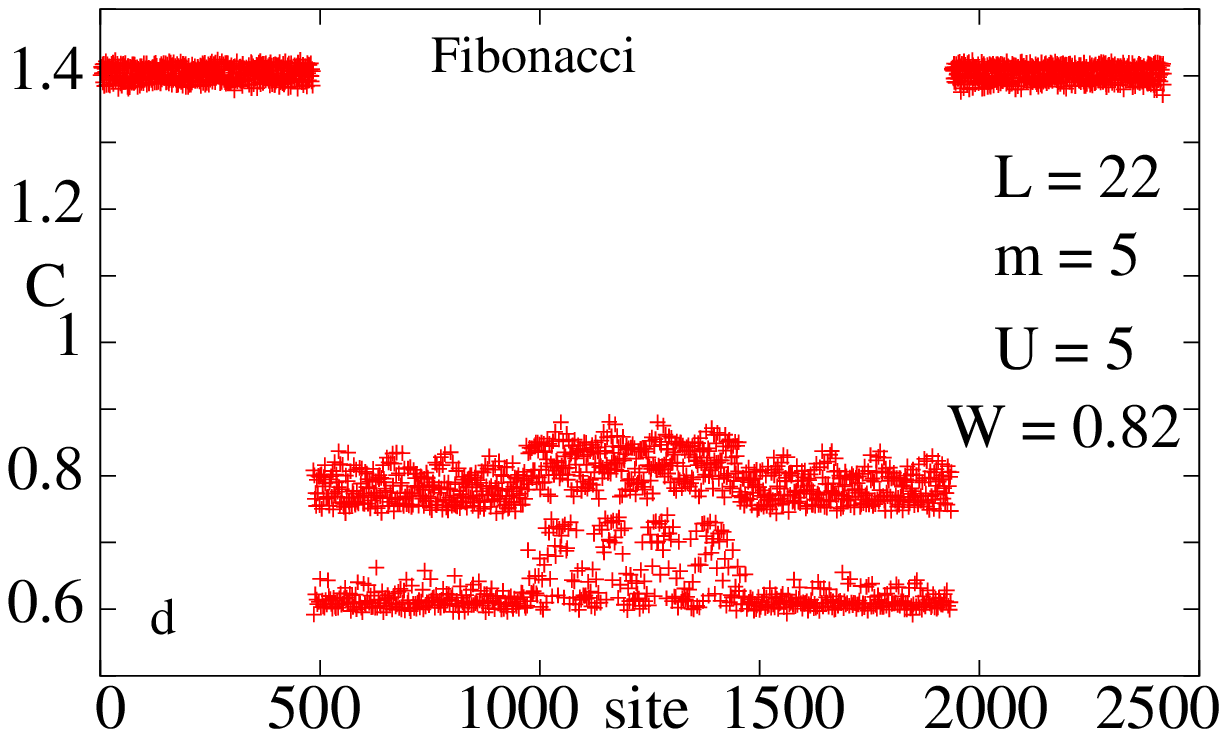}} \\
    \end{tabular}
\caption{Spin and charge profiles for Fibonacci type of disorder.
a. Spin at each site for $U = 5$ and $W = 0.6$. 
b. Charge at each site for $U = 5$ and $W = 0.6$.
c. Spin at each site for $U = 5$ and $W = 0.82$.
d. Charge at each site for $U = 5$ and $W = 0.82$. }
\label{fig;Spin_ChargeOrder_Fibonacci}
\end{center}
\end{figure}

\subsection{Results and Analysis}

Our results in this paper corresponds to $U=5$ in presence of disorder.
We have chosen this value as representative of the region where the 
clean system is a Mott insulator. As we increase the disorder strength, 
the gap at half filling closes and the system becomes metallic. 
There is also a gradual delocalization of the wavefunction as the disorder strength 
is increased. 
The spin order parameter becomes highly dispersed in each of the planes 
as the metallic phase is approached. The peak conductivity values are roughly the 
same for both types of correlated disorder considered in this work. 

%\subsection{Analysis of our results} 

Fig 1a shows the phenomenon of the gap at half filling, due to correlation,
closing due to the presence of disorder. Disorder introduces states at 
half filling. The concentration of $B$ sites is more for the Thue Morse 
sequence compared to the Fibonacci sequence. Hence the Thue Morse sequence is 
able to kill the gap more quickly than the Fibonacci sequence. Thus we see 
that while the Thue Morse sequence kills the
gap to a value less than the finite size spacing at roughly $W = 0.6$, the 
Fibonacci sequence is able to achieve the same 
reduction in gap at around $W = 0.82$. 
 
Fig 1b looks at the change in the extent of localization of the wavefunction
 as we increase the disorder through the Inverse Participation Ratio(IPR). High
IPR values indicate higher localization while low IPR values indicate delocalized
wavefunctions. The IPR decreases monotonically with increasing disorder, thus 
indicating that the wave functions become more delocalized. The calculation has 
been done for both Thue Morse and Fibonacci sequence of disorder. The 
delocalization process is more rapid for the Thue Morse type and is more gradual 
for the Fibonacci sequence. We have not been able to go beyond the transition point 
due to the onset of a glassy phase which makes it very difficult to calculate the 
correct configuration accurately.

\begin{figure}
\begin{center}
   \begin{tabular}{cc}
      \resizebox{39.5mm}{!}{\includegraphics{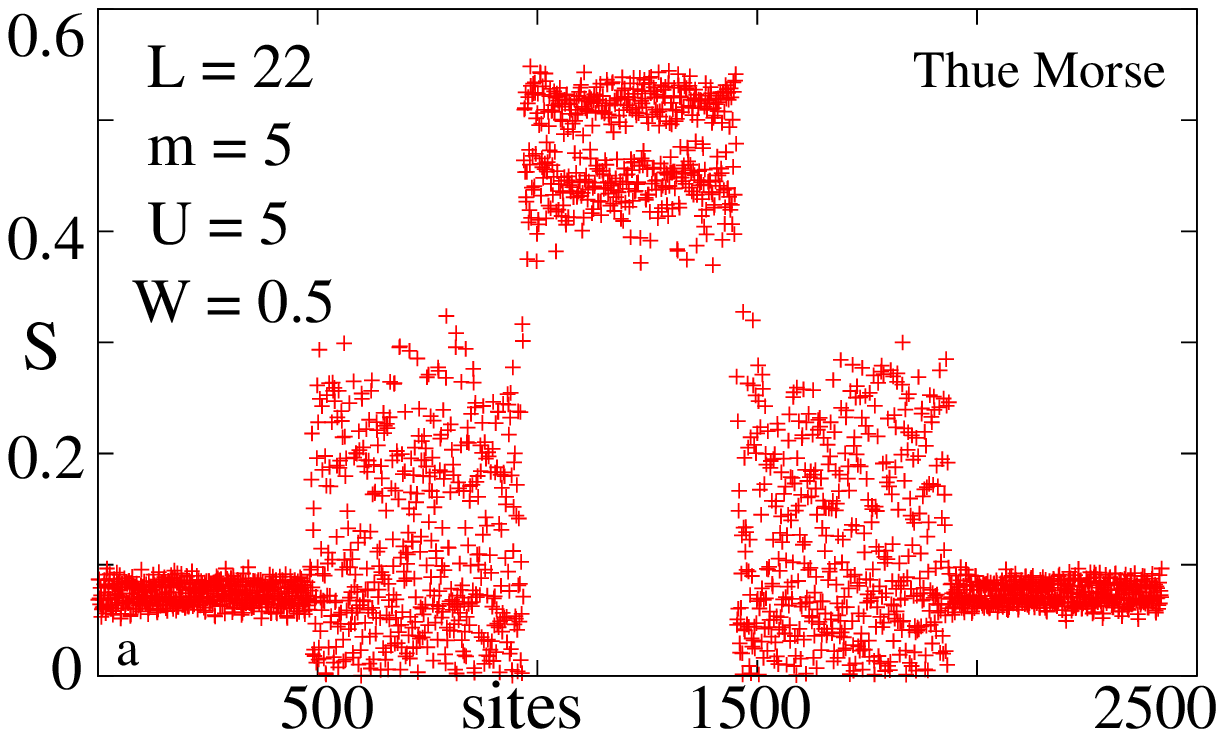}} &
      \resizebox{39.5mm}{!}{\includegraphics{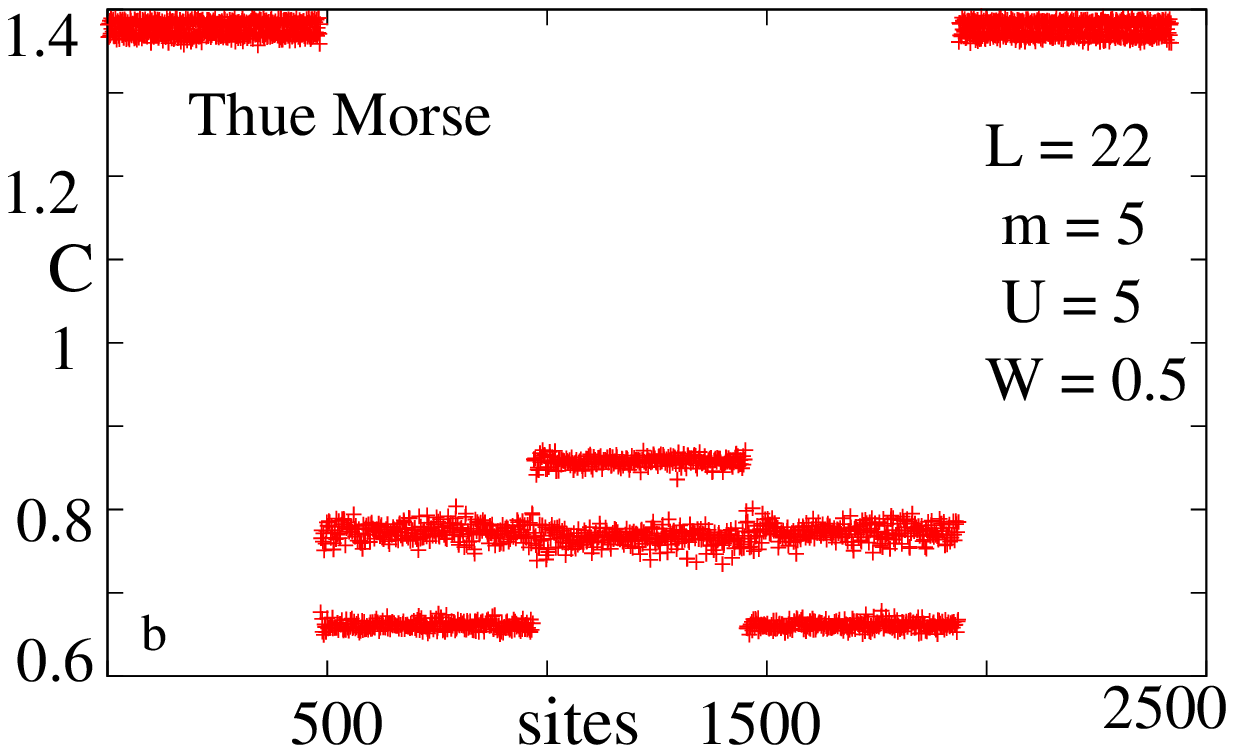}} \\
      \resizebox{39.5mm}{!}{\includegraphics{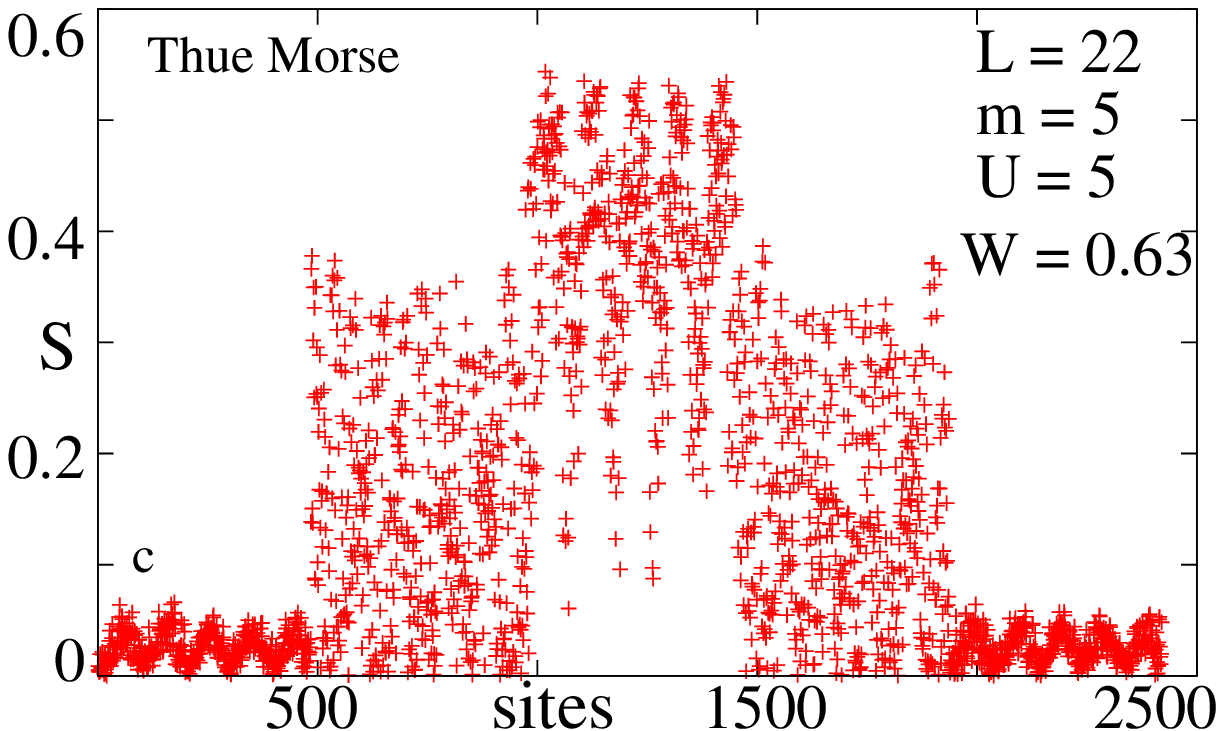}} &
      \resizebox{39.5mm}{!}{\includegraphics{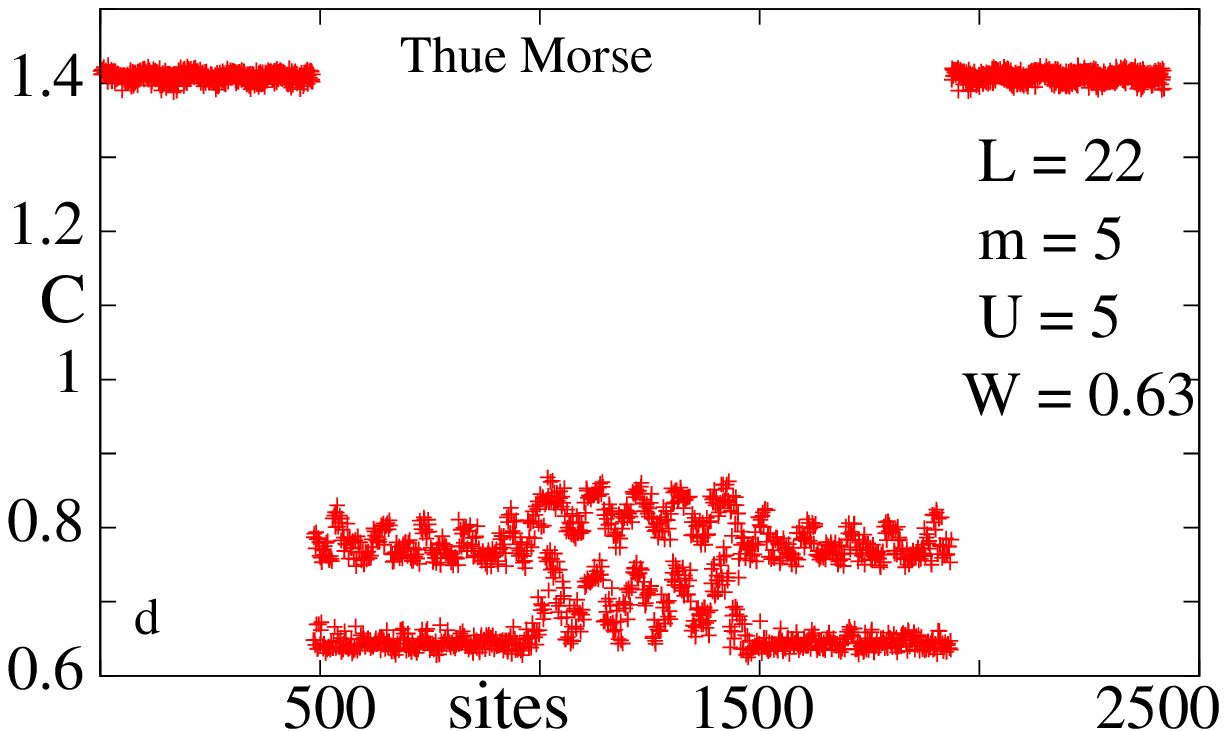}} \\
    \end{tabular}
\caption{Spin and charge profiles for Thue Morse disorder. a. Spin at each 
site for $U = 5$ and $W = 0.5$. b. Charge at each site for $U = 5$ and $W = 0.5$
c. Spin at each site for $U = 5$ and $W = 0.63$, d. Charge at each site for 
$U = 5$ and $W = 0.63$}
\label{fig;Spin_ChargeOrder_Thue_Morse}
\end{center}
\end{figure}

From our previous studies \cite{Gupta2} we know that for $m=5$ which 
means 3 intervening Mott layers, there is a gap at half filling for 
$U=5$ accompanied by the induction of anti ferromagnetic ordering.
The Mott layers induces an anti ferromagnetic
ordering even in the metal planes. The ordering is non uniform as one 
moves along the z-axis with the maximum value at the central plane.   

Fig 2 and Fig 3 shows the spin and total charge at each site for Fibonacci and 
Thue Morse type of disorder respectively, for $L = 22$, $m = 5$, $U = 5$. Thus 
in the figures, the first $L^2$ sites correspond to the lowest metallic plane, 
the next $L^2$ sites correspond to the adjacent Mott plane, the next $L^2$ to the 
central layer Mott plane and so on. There is obvious symmetry about the central plane 
in the spin and charge order parameters.  

Fig 2a and 2c show the spin profile for the insulating($W = 0.6$) and metallic
($W = 0.82$) regime respectively, while Figs 2b and 2d show the charge profile 
for the same values of $W$. Figs 3a and 3c shows the corresponding spin profile 
for the insulating($W = 0.5$) and metallic ($W = 0.63$) regime respectively, for 
the Thue Morse sequence, while Figs 3b and 3d show the charge profile for the 
same values of $W$. 
 
We find disorder increases the total charge buildup in 
the metallic layers. The charge depletion from the 
Mott layers just adjacent to the metallic planes and the 
the charge accumulation in the metallic planes increases with increasing $W$.  
The spin profile becomes totally dispersed for the metallic phase in contrast 
to a less dispersed profile in the insulating phase. 

From the Kubo conductivity plots shown in Figs 4a and 4b we find that
as we increase disorder for fixed $U = 5$, the system shows a metal 
insulator transition(MIT). The conductivity value changes non monotonically 
with increasing disorder in the metallic phase. 
 
\begin{figure}
\begin{center}
   \begin{tabular}{cc}
      \resizebox{39.5mm}{!}{\includegraphics{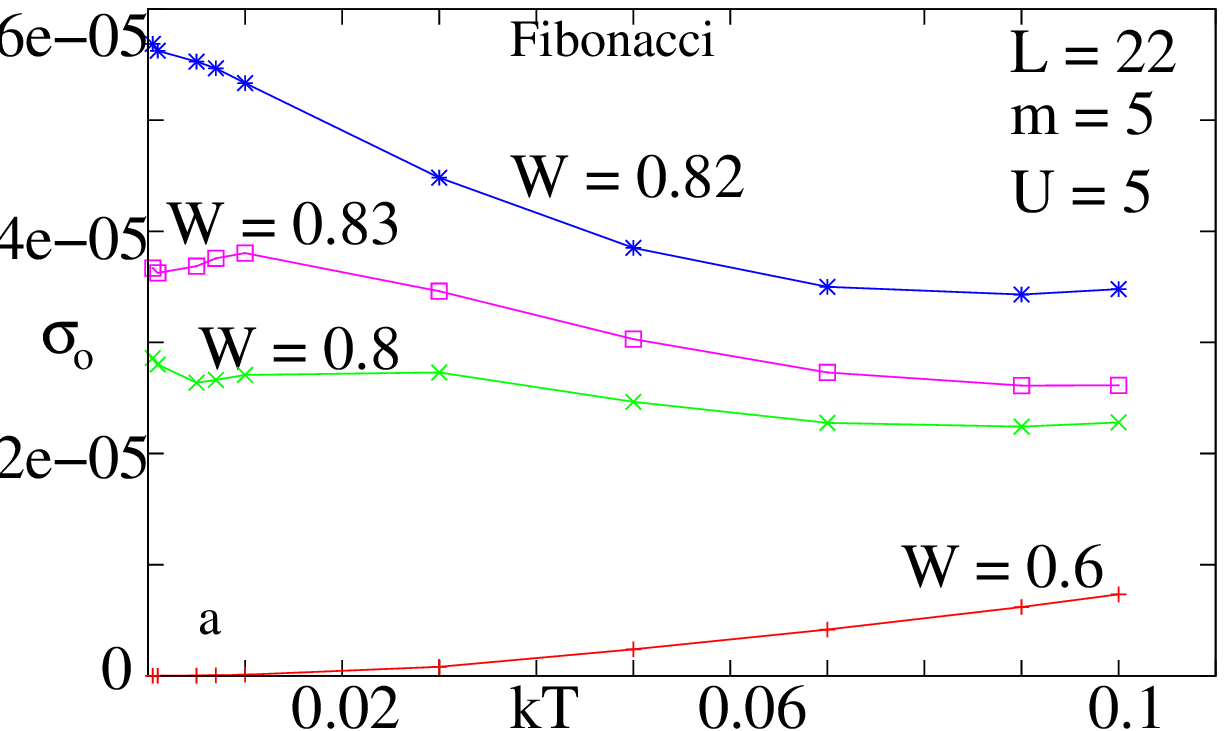}} &
      \resizebox{39.5mm}{!}{\includegraphics{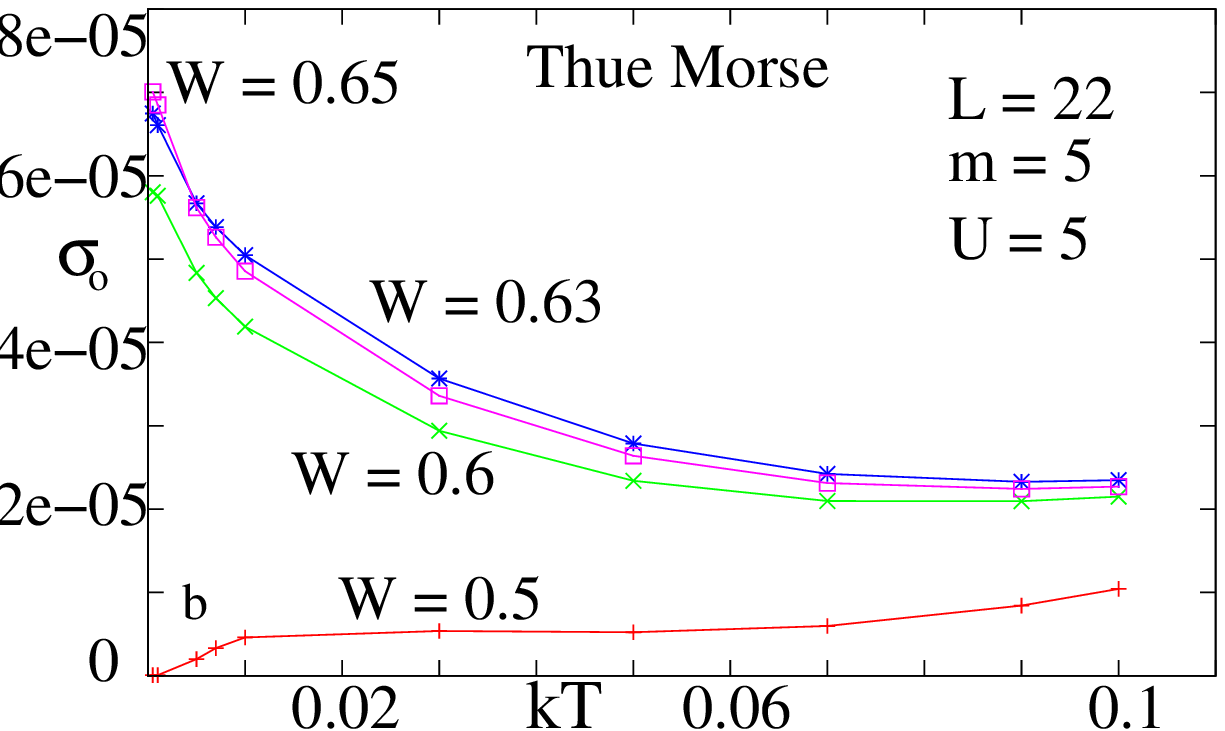}} \\
    \end{tabular}
\caption{a. Insulator-Metal-Insulator transition observed with increasing disorder  
strength of the correlated Fibonacci type.}
\label{fig;MIT}
\end{center}
\end{figure}

\section{Conclusion}
A metal-disordered Mott insulator-metal interface heterostructure is studied. A 
disorder induced metal insulator transition is observed, over a small range of 
disorder values, in the range where the clean system is gapped. 

\vskip .1in
Sanjay Gupta would like to thank the DST for providing the grant under the
project "Electronic states and transport in mesoscopic/nanoscopic systems".

\end{document}